\documentclass{cmmp}
\usepackage{epsfig}

\let\realverbatim=\verbatim
\let\realendverbatim=\endverbatim
\renewcommand\verbatim{\par\addvspace{6pt plus 2pt minus 1pt}\realverbatim}
\renewcommand\endverbatim{\realendverbatim\addvspace{6pt plus 2pt minus 1pt}}

\renewcommand\thesection{\arabic{section}}

\def\IR{\relax{\rm I\kern-.18em R}}
 
\begin{document}
\thispagestyle{empty}
\setcounter{chapter}{1}

 
\rightline{\vbox{\baselineskip12pt
\hbox{CITUSC/02-018}
\hbox{USC-02/03} 
\hbox{\tt hep-th/0205201} }}

\bigskip
\bigskip
\bigskip
\bigskip
\bigskip
\centerline{\Large\bf Gauged Supergravity and Holographic Field Theory
\footnote{ This work was supported in part by funds  
provided by the DOE under grant number DE-FG03-84ER-40168. }}
\bigskip
\bigskip
\bigskip
\bigskip
\centerline{{\bf Nicholas P. Warner}
\footnote{e-mail address:   warner@physics.usc.edu.  }}
\bigskip
\bigskip
\centerline{{\it Department of Physics and Astronomy}} 
\centerline{{\it and}}
\centerline{{\it CIT-USC Center for
Theoretical Physics}}
\centerline{{\it University of Southern California}} 
\centerline{{\it Los Angeles, CA
90089-0484, USA}} 
\bigskip
\bigskip 

\bigskip
\bigskip
\bigskip
\centerline{\bf Abstract}
\bigskip
\noindent
This is a slightly expanded version of my talk at 
{\it Future Perspectives in Theoretical Physics and Cosmology,} 
Stephen Hawking's 60th Birthday Worshop.  I describe some 
of the issues that were important in gauged supergravity 
in the 1980's and how these, and related issues have once again
become important in the study of holographic field theories.  
 \newpage
\setcounter{page}{1}

\chapter*{Gauged Supergravity and Holographic Field Theory}

\section{Gauged supergravity and a thesis project}

I became one of Stephen's students a little over 20 years ago.  At
the time, Stephen was formulating and developing many of his 
ideas of Euclidean quantum gravity, and was also greatly 
interested in, and supportive of, other approaches to quantum gravity.
Most particularly, he was an enthusiastic advocate of 
supergravity theories.  Supergravity theories grew out of 
particle physics\footnote{I would like to apologize in advance
to the  large number of people who made huge contributions 
to the development of supergravity and string theory, but my allotted space 
makes it impossible to adequately and correctly reference them all.}
and were not the standard fare of Relativists.   It is a testament to
Stephen's broad interests and the atmosphere in the Cambridge Relativity 
Group that the lines between these disciplines were completely blurred 
and that it was possible -- indeed encouraged --  for me as a student to 
move from my first work in relativity and Euclidean quantum gravity to 
the study of symmetry breaking in gauged supergravity.   
I remain very grateful to Stephen and to the Relativity Group for this
opportunity and, most particularly, for the  excitement, interest and 
enthusiasm with which they  pursued the  ``Grand Enterprise'' of looking 
for a viable quantum theory of gravity.

The primary problem of quantum gravity is the infinite number of
different kinds of divergence in the perturbation theory of pure quantum 
gravity.  It was well known that divergences caused by fermions generically
have the opposite sign to similar divergences caused by bosons.  In a 
supersymmetric theory, bosons are paired with fermions in just such a
manner that these divergences have the same structure and tend to cancel 
one another in the  quantum theory.   In a supergravity theory, the graviton
is paired with one, or more, fermionic partners, called gravitini.  The number 
of gravitini, ${\cal N}$, denotes the amount of (extended) supersymmetry.  
The more supersymmetry the larger the spectrum of the theory:  
${\cal N} =1$ supergravity has simply a graviton and one gravitino, and the 
maximal supergravity in four dimensions has  ${\cal N} =8$, with a spectrum of
one graviton, eight gravitini, $28$ vector bosons, $56$ spin-${1\over 2}$ 
particles and $70$ scalar fields.  One cannot go beyond ${\cal N}=8$ since
the spectrum would then have to include particles whose spin is higher
than that of the graviton and, as far as we know, non-stringy higher spin 
theories appear to be inconsistent.
The important point is that, in general, the more supersymmetry there is,
the more the divergences are cancelled.  The crucial question of the
early 1980's was whether ${\cal N}=8$ supergravity was finite.
It was ultimately shown that even maximal supergravity very probably had 
divergences at higher orders \cite{Kallosh:1980fi,Howe:1981xy},   
but despite this difficulty,  it was
clear that supergravity was a very important step in the right direction and
that it had to be part of the right answer.   Subsequent history has born this 
prejudice out in that supergravity is the low energy limit of string theory,
which {\it is}  (as far as we can tell) a finite quantum theory of gravity.

In the early 1980's, when we were still trying to turn supergravity into the 
{\it Theory of Everything}, there was another important issue:  Maximal 
supergravities  contained vector bosons, but in the original formulations 
all these vector bosons in these theories  were Abelian.  To describe the real 
world one needs non-abelian gauge symmetry, and many people  painstakingly
constructed the so-called gauged supergravity theories, that is, theories
with non-abelian gauge symmetry mediated by the vector bosons.  This  
work culminated  in the eventual  construction of gauged, maximal 
(${\cal N}\! =\! 8$)  supergravity theory \cite{deWit:1982ig}.  This theory 
also had the maximal possible gauge group  of $SO(8)$. 

It is a remarkable and rather attractive feature of these gauged
supergravity theories that the joint requirements of gauge symmetry 
and supersymmetry also requires a
non-linear potential in the scalar sector.  Moreover, the more the supersymmetry,
the more rigid the potential, and indeed the potential is completely
fixed in the maximal theory.  Thus, the maximal gauged theory
determines its own symmetry breaking structure completely; there are no
choices and no arbitrary parameters to be fixed.

In the early days of supergravity the focus, for obvious reasons, was 
primarily upon supergravity in four space-time dimensions.  However,
it became very important to study supergravity in  every possible 
dimension.  There are several reasons for this (some of
which will be described later), but one of the primary reasons was that the
higher dimensional maximal theories are generically simpler,
and thus easier to construct.  Indeed the lower dimensional
maximal theories were often first constructed by dimensionally
reducing the higher dimensional theories (see, for example
\cite{Cremmer:1978km,Cremmer:1979up}).    
Here I will be concerned mainly with the gauged
maximal supergravities in four dimensions and in five dimensions;
both of these theories have ${\cal N}=8$ supersymmetry.   The latter
theory has  a spectrum of  one graviton, eight gravitini, $15$ vector bosons, 
$12$ tensor gauge fields,  $48$ ``spin-${1\over 2}$''  particles
and $42$ scalar fields; the  gauge group is $SO(6)$.
The five-dimensional theory was also 
the last gauged maximal supergravity to be constructed 
\cite{Gunaydin:1984qu,Gunaydin:1985cu, Pernici:ju}.
This was, to some extent, because of technical issues, but also
because it seemed of the least phenomenological interest.  
It is thus a wonderful irony that this situation is now reversed.

As I walked home with him one wet winter's day, Stephen suggested
that I should study the four-dimensional, maximal gauged ${\cal N}=8$ 
theory and  try to understand its symmetry breaking
structure.  This was to be the last project of my PhD.  At the time,
neither I nor Stephen remotely suspected that, rather than finding its most
important application directly within quantum gravity, this work would prove 
important 20 years later in  determining part of the phase diagram 
of strongly coupled, large $N$, ${\cal N}=4$ Yang-Mills theory:  A theory that 
is a distant relative of QCD, which describes the force that underpins the strong 
nuclear interaction.  It is  the puposes of this talk to outline how all of
this came about, and describe how the ideas of gravity and supergravity 
can be used to give beautiful and remarkable insights into field theory via 
the idea holography on branes.

\section{The up's and down's of maximal gauged supergravity}

In any supergravity theory, or string theory, one wants to find
interesting, and hopefully viable, ground states for the theory.
In gauged supergravity one is thus naturally led to study the
scalar fields and their potential. Both the
four and five dimensional maximal gauged supergravity theories have
a completely determined scalar  potential,  and these
potentials have somewhat similar structural features.  Thus, while
my remarks in this section will be directed toward the four-dimensional theory,
many of the results have five-dimensional parallels that will
be important later.

\subsection{The universe is not anti-de Sitter}

It is a fundamental property of the superalgebra underlying
gauged supergravity that if a ground state preserves
supersymmetry, and that supersymmetry transforms under
a non-abelian gauge symmetry, then the ground state must
necessarily have a negative cosmological constant.  This 
cosmological constant, $\Lambda$, is  of order $-g^2$ in Planck units,
where $g$ is the gauge coupling constant in the supergravity.

This fact means that the easiest ({\it i.e.} most symmetric)
ground states to find will generically be associated with
anti-de Sitter (AdS) space with Planck scale $\Lambda$.  In particular,
this includes  the maximally supersymmetric ground state,
in which all the scalars vanish.  An optimist
would note that ${\cal N}=0,1$ or ${\cal N}=2$ supersymmetric
vacua with zero cosmological constant are still allowed since
such theories have either no gauge symmetry action, or  an (abelian) 
$U(1)$ action on the residual supersymmetry generators.  Thus
a phenomenologically interesting ground state is not excluded by
the AdS superalgebra.  But there is still  the problem of explaining the transition 
from the maximally supersymmetric AdS vacuum to the (hopefully) 
flat vacuum.

\subsection{Stability of ground states}
 
There was also the more immediate problem that all the ground
states appeared to be pathologically unstable, even the maximally
supersymmetric  ground state. 
\begin{figure} 
\epsfig{figure=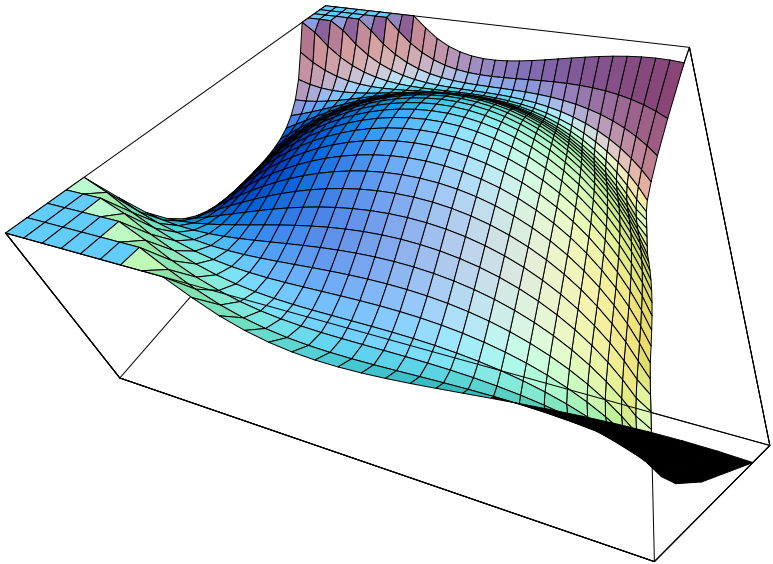,width=4in}
\vspace{5pt}
\caption{A generic contour map of a two-dimensional section
through the maximally symmetric critical point of a supergravity potential.
The hump in the middle is the maximally supersymmetric local maximum.}
\label{Vplotfig}
\end{figure}
Figure \ref{Vplotfig} shows a contour map of a typical two-parameter 
section through the maximally supersymmetic critical point.  This
point is a local {\it maximum} of the potential, and is thus naively unstable.

It turns out that any supersymmetric critical point is completely
classically (and semi-classically) stable \cite{Gibbons:aq}.  One can prove this
by establishing an AdS positive mass theorem.  The key physical
insight is that because the potential is Planck scale, the gravitational
back-reaction is very strong.  Indeed Breitenlohner and Freedman
showed very generally that in AdS space, a scalar field was perturbatively
stable so long as its mass-squared is not ``too negative''  \cite{Breitenlohner:jf}.  
This bound in an anti-de Sitter space, $AdS_d$,  of
dimension $d$ and radius $R$ is  \cite{Breitenlohner:jf,Townsend:iu}:
\begin{equation}
m^2 ~\ge~ - {(d-1)^2 \over 4\, R^2} \,.
\label{BFbound}
\end{equation}
 \subsection{The standard model and chiral fermions}

The maximal possible gauge symmetry of maximal supergravity is
$SO(8)$, and this group is not large enough to contain the gauge
groups of the standard model: $SU(3) \times SU(2) \times U(1)$.
There were suggestions that the physical gauge groups might emerge
through some composite mechanism, or via a large $SU(8)$ 
local composite symmetry of the supergravity theory.  These
ideas have remained just that, and were not given substance in
four dimensions.

The other problem was that the fermions come in real 
representations of the $SO(8)$ gauge symmetry, and
even if one could break to $SU(3) \times U(1)$ (hoping
that $SU(2)$ would emerge somehow at low energy)
then the fermions would still be in non-chiral (real)
representations of these groups.   Thus $SO(8)$ seemed
to be fairly hopeless in terms of phenomenology 
(in spite of some remarkable numerology that originated
with Gell-Mann:  See  \cite{Nicolai:1985hs} for details).

Thus the ``real-world'' possibilities were very remote for  this 
most  divergence-free theory of all supergravities.  Moreover, by this
time, a consensus emerged that this theory was also most probably  
not  finite.  Thus the gauged maximal supergravity slowly faded
from notice and interest.

\section{Exploring higher dimensions}

Parallel to the development of maximal supergravity
in four-dimensions was the development and construction
of supergravity in higher dimensions.  As I remarked
earlier, the ungauged maximal theories were first obtained by
``trivial'' ({\it i.e.} on tori) dimensional reduction of the eleven-dimensional
supergravity.  It was one of the major industries of the
early 1980's to see what low-dimensional theories might
be obtained by compactifying higher-dimensional maximal
theories on other manifolds.   

Initially the focus was upon Kaluza-Klein methods, in which gauge
symmetries in lower dimensions emerged via isometries 
of the compactifying manifold. As with gauged supergravity, there 
were huge phenomenological problems with this.  First, isometries on the
compactifying manifold tended to lead to AdS space-times.   Furthermore,
it was  finally shown  \cite{Witten:me,Wetterich:1983ye,Witten:1985xc}
 that one could never get chiral fermions
though Kaluza-Klein without having chiral fermions {\it ab initio}.

In the mid 1980's string theory emerged from a long hibernation, and
with the invention of the heterotic string it appeared that one could
finally get a finite theory of quantum gravity with large enough gauge
groups,  chiral fermions and no anomalies.  The focus thus turned to the 
compactificaton
of string theory, some of  which entailed finding compactifications
of the corresponding low-energy supergravity theories (often coupled
to supersymmetric matter).  

\subsection{Sphere compactifications}

One particular focus of supergravity was the compactification of 
eleven-dimensional supergravity on $S^7$ down to $AdS_4$.  
It was believed, and indeed  subsequently proven,
that the low energy sector ({\it i.e.} essentially the lowest Fourier modes 
on $S^7$) of this theory is, in fact, maximal gauged 
${\cal N}=8$  supergravity in four dimensions.  The complete set of
Fourier modes of the  $S^7$ compactification
would thus extend maximal gauged  ${\cal N}=8$ supergravity by infinite
towers of massive states.

A variant of this idea is of particular relevance today.  There is a maximal
supergravity theory, called IIB supergravity, in ten dimensions that has 
two chiral fermions \cite{Schwarz:qr} (and so is not a trivial dimensional 
reduction of  eleven-dimensional supergravity). 
This theory is a low-energy limit of the  ten-dimensional, IIB superstring.  
This supergravity, and the corresponding string theory theory have a 
compactification on $S^5$ down to five dimensional $AdS_5$, and this 
almost certainly  (though it was never fully proven) yields, in the massless 
sector, the maximal  gauged ${\cal N}=8$  theory in five dimensions.
 
This string back-ground also arises in another very interesting manner.

\subsection{Brane backgrounds}

\begin{figure} 
\epsfig{figure=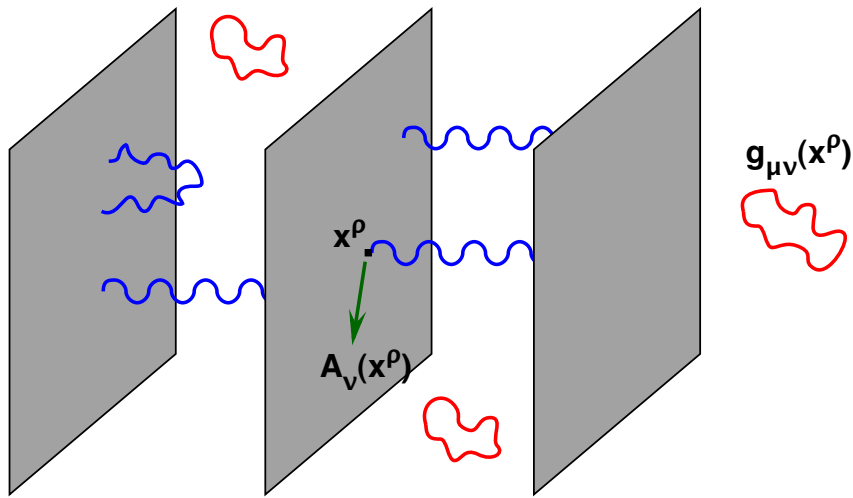,width=4.5in}
\vspace{4.5pt}
\caption{Open strings can end on the $D$-branes, while only closed strings 
appear in the bulk. The lowest modes of the closed string describe the graviton 
supermultiplet, while the lowest modes of the open strings describe 
vector supermultiplets localized on the $D$-branes.}
\label{branes}
\end{figure}

A very important class of stringy backgrounds are $p$-branes, which
are the higher dimensional 
analogues of the extreme Reissner-Nordstr\"om black holes.  That is,
they are $(p,1)$-dimensional objects that minimally couple
to a $(p+1)$-form  gauge potential.  
If the gauge field is a Ramond-Ramond (RR)  field of the string then they
are called $Dp$-branes,  and most significantly, in a closed string theory
there can also be open strings that end  on the $D$-branes  
\cite{Polchinski:1995mt}.  The lowest modes of the closed string generically 
describe the graviton supermultiplet, and the lowest modes of the open string 
generically describe a vector supermultiplet. The directions of the oscillations of
the string give rise to the polarization tensors (see figure \ref{branes}).

The consequence of this is that a closed string in the presence of
$Dp$-branes will induce a Yang-Mills theory on the $D$-branes.
If there are $N$ coincident $D$-branes then the Yang-Mills theory has
an $SU(N)$ gauge group, and the amount of supersymmetry on the
brane is half of that of the bulk theory.  Thus $IIB$ superstrings induce 
${\cal N}=4 $ supersymmetric, $SU(N)$ Yang-Mills theory on a stack
of $N$ coincident $D3$-branes (whose world-volume is four
dimensional).   This particular Yang-Mills theory is also a conformal
field theory (CFT), even as a strongly coupled quantum theory.

\begin{figure} 
\epsfig{figure=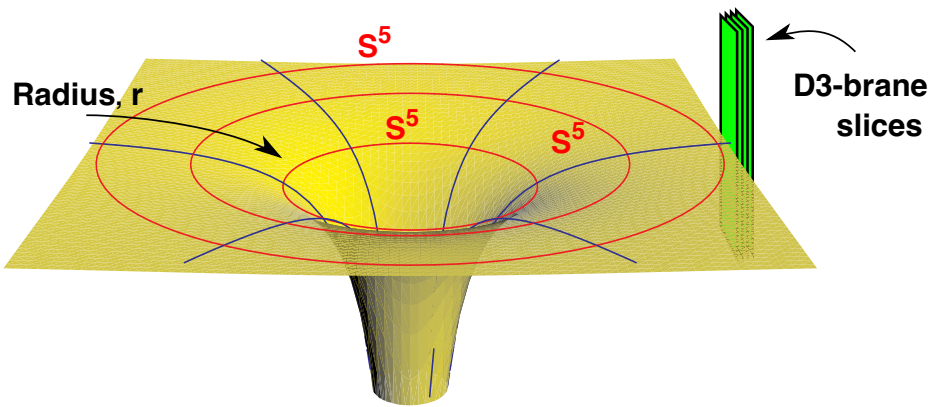,width=5in}
\vspace{5pt}
\caption{The near-brane limit of $D3$-branes: The circles denote the
$S^5$'s around the branes in the core of the solution, while
the radial coordinate and the slices parallel to the $D3$-branes 
combine to make an $AdS_5$.}
\label{nearbrane}
\end{figure}

The extreme Reissner-Nordstr\"om black hole has a ``near-horizon''
limit that is essentially an infinitely long throat described 
by $AdS_2 \times S^2$.  In the same manner, the near-brane limit of
the $D3$-brane background is $AdS_{5} \times S^{5}$.
The metric in this near-brane limit may be written:
\begin{equation}
ds^2_{10} ~=~ e^{2\, A(r)} \, \Big( \eta_{\mu\nu}\, dx^\mu \, dx^\nu \Big) ~+~
dr^2 ~+~ ds_5^2 \,,
\label{tenmet}
\end{equation}
where $A(r) =  r/L$ and $ds_5^2$ is the metric of an $S^5$ of radius $L$.
The  factor $\eta_{\mu\nu}\, dx^\mu \, dx^\nu$ is the flat metric on
the $\IR^{3,1}$ slices parallel to the $D3$-branes.  The first two terms
in (\ref{tenmet})  combine to make the $AdS_5$ metric in which 
$L = (4\pi g_s \alpha' N)^{1 \over 4}$ is the AdS radius.  Here 
$N$ is the number of $D3$ branes,   $g_s$ is the string coupling and
$\alpha'$ is the string tension parameter.
Thus, as depicted in Figure \ref{nearbrane}, this maximally supersymmetric
 background of the $IIB$ superstring emerges as an infinitely long throat
as one tries to approach the horizon of the $D3$-brane background.

The next leap forward was made by Maldacena \cite{Maldacena:1997re} 
in the  late 1990's, and
it was to conjecture a deep relationship between the physics of
the closed IIB superstring theory (and hence IIB supergravity) in the
``near-brane'' background and the
physics of the Yang-Mills theory on the $D3$-branes.

\section{Holographic field theory and the AdS/CFT correspondence}

As discussed by other speakers at this meeting, the basic idea in 
holographic field theory is  that, in the presence of gravity,  
physics in a region of space-time can be encoded
on a lower-dimensional surface surrounding that region. In particular,
the quantum properties of  matter that has formed a black hole
can be (and have been) holographically encoded on its horizon
\cite{'tHooft:gx,Susskind:1994vu}.  
The Maldacena Conjecture\footnote{By now it is sufficiently widely believed
and tested that we should probably call it a theory or principle.} makes
precise computational proposals for several such  holographic encodings.
 One of these states that there is a precise quantum duality between IIB 
superstring theory in an $AdS_5 \times S^5$ background and the CFT
on $D3$-branes, that is, ${\cal N}=4$ Yang-Mills theory.  The
holographic principle was originally intended as a way of studying gravity
by looking at a holographic field theory on a surface, but here
I will be using the holographic principle in reverse to
derive field theory results on the brane using $IIB$ string theory, or
more precisely supergravity.  
 
The heart of this AdS/CFT correspondence is to realize that a gauge
invariant operator, ${\cal O}(x)$, in the Yang-Mills theory on the brane
will act as a source for closed strings in the bulk.     In particular,
conserved currents on the brane must couple to local gauge fields
in the string background.  For example, the energy-momentum tensor
of the Yang-Mills theory couples to (or is dual to) the graviton in the
string theory, and the currents of the global $SO(6)$ $R$-symmetry of 
Yang-Mills are dual to the vector bosons of the gauged supergravity.  
This idea extends, via supersymmetry, to a complete correspondence
of fields in the Yang-Mills energy-momentum supermultiplet with the
 fields in the graviton supermultiplet.
The  latter  are precisely the massless fields that constitute the spectrum
of maximal gauged ${\cal N}=8$ supergravity in five dimensions.   
In particular, the $42$ scalar fields of the gauged  supergravity
are dual to bilinears of fundamental Yang-Mills fields, or roughly
the Yang-Mills gauge coupling, $\theta$-angle, and all the
mass terms for the Yang-Mills fermions and scalar fields.

More generally, if one introduces a generating function for correlation
 functions of gauge invariant Yang-Mills operators, then
the AdS/CFT correspondence  states that 
\cite{Maldacena:1997re,Gubser:1998bc,Witten:1998qj}:
\begin{eqnarray}
\left< \exp\left( - \int \varphi^{(0)}_j~{\cal O}_j (x_j) ~d^4x \right) \right> 
\bigg|_{\rm brane}  &= & {\cal Z}_{\rm string} [\varphi_k] \cr
  &\to & {\cal Z}_{\rm supergravity} [\varphi_k]  \cr 
  &\to &   \exp\left( - {\cal S} [ \varphi_k]   \right) \,.
\label{corresp}
\end{eqnarray}
In this equation the left-hand side is the generating function with operators
integrated against arbitrary density functions, $\varphi^{(0)}_j$, on the branes.
The function ${\cal Z}_{\rm string} [\varphi_k] $ is the string path integral evaluated
in the $AdS$ background fields, $ \varphi_k $, that satisfy boundary conditions
\begin{equation}
 \varphi_k (x^\mu,r)   ~\to~ \varphi^{(0)}_j(x^\mu) \, \quad {\rm as} \quad  
r \to \infty \,,
\label{bconds}
\end{equation}
where $r$ is $AdS_5$ radial coordinate transverse to the branes,
and is defined in (\ref{tenmet}).
The first limit in (\ref{corresp}) reduces the string theory to its
supergravity limit, and this may be done by a combination
of taking the string tension to infinity ($\alpha' \to 0$) and 
sending $g_s N \to \infty$.  The second limit
in  (\ref{corresp}) is obtained by making the saddle-point
approximation ($g_s \to 0$) of the supergravity path integral, 
${\cal Z}_{\rm supergravity} [\varphi_k]$, and thus one
simply evaluates the supergravity action on the classical 
supergravity solution that satisfies the boundary conditions
(\ref{bconds}).
The strongest form of the correspondence is, of course, the
string theory expression, but, in practice, it is much easier 
to work with supergravity actions, and it is in this
context that most (but not all) of the testing has been done.

The ``bottom line'' is that the classical action of $IIB$
supergravity in ten dimensions  should  holographically capture 
the large-$N$ limit of strongly coupled quantum Yang-Mills theory.
Moreover, if one restricts to the operators of the energy-momentum
tensor supermultiplet (mass insertions, gauge coupling and  $\theta$
angle) then the behaviour of this large-$N$ strongly quantum theory on
the brane, under such perturbing operators, should be captured
entirely by gauged, ${\cal N}=8$ supergravity in five dimensions.

\section{Bulk gravity and brane renormalization: Where are the branes? }
 
{}From the gravitational perspective the branes must be at the core of
the solution, that is, at the bottom of the infinitely long throat.  
Therefore,   the boundary condition (\ref{bconds}) as $r \to \infty$ seems 
very counterintuitive: The branes are at  $r= -\infty$ and not at $r= +\infty$.
However, locating the branes is not so simple in the holographic
perspective:  In holography one considers the field theory on any
of the $(3+1)$-dimensional Poincar\'e invariant slice of the space-time.
The idea is that one is then ``sampling'' the field  theory on a brane located
 at some radius,  $r$, in the metric (\ref{tenmet}), and the choice
of the radius represents a choice of the renormalization scale in the field
theory.  I want to illustrate this in several ways since this idea is
central to some of the more remarkable tests and applications of
the AdS/CFT correspondence.

First, one should recall that  $AdS_5$ is invariant under $SO(4,2)$,
and that this is the conformal group in $(3+1)$ dimensions.  If one takes
a general $D3$-brane slice of (\ref{tenmet}) then fixing a finite value
of $r$ breaks $SO(4,2)$ down to the Poincar\'e invariance on the slice, 
and in particular, scale invariance has been broken.   However, the slice 
at $r = \infty$ is special:  It is fixed under the action of $SO(4,2)$, and so 
the field theory induced on this slice is conformally invariant.  Thus we 
associate  the brane at infinity with a conformal, ultra-violet fixed point of the 
brane field theory, and a brane at finite  $r$ with the theory
at some cut-off scale set by $r$.   Indeed, to implement  many
computations using (\ref{corresp}) one has to regulate the Green functions,
and this can be done explicitly by setting  
$r  = 1/\epsilon$ and sending $\epsilon \to 0$.   (See, for example 
\cite{Gubser:1998bc,Witten:1998qj,Henningson:1998gx}).  As one might expect, 
the true physical scale on the brane is set by the cosmological scale factor, 
$e^{A(r)}$, in  (\ref{tenmet}), and not so much by $r$ itself.

\subsection{Coarse-graining and red-shifts}
\begin{figure}[h]
\epsfig{figure=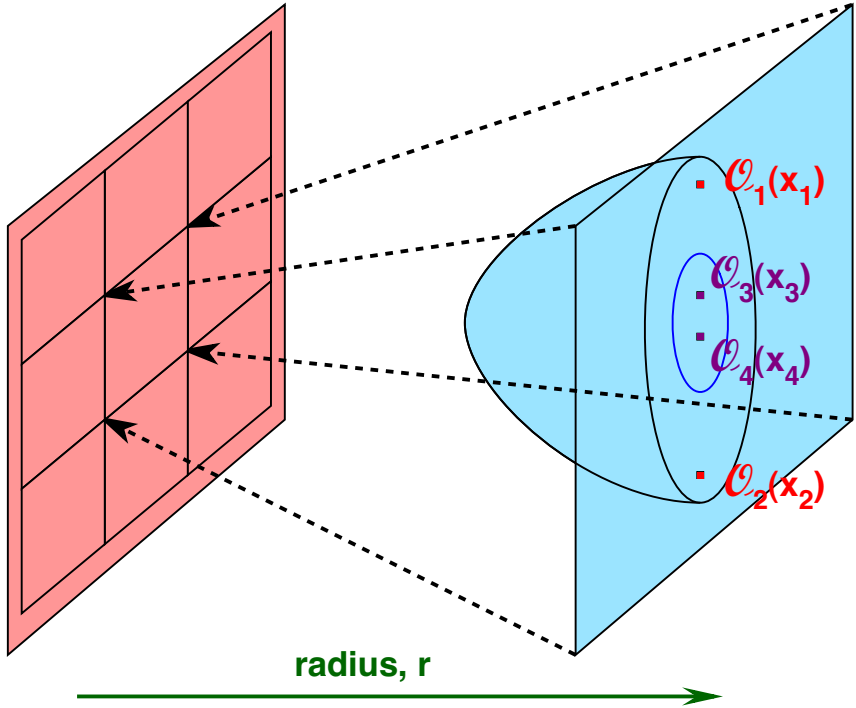,width=4.5in}
\vspace{4.5pt}
\caption{The UV-IR duality of holgraphy:  Large-distance behaviour on the 
brane depends upon details at small radii, $r$, while short-distance
behaviour depends only upon large $r$.  Alternatively, gravitational red-shifts
generate Wilsonian coarse graining on the brane. }
\label{rgaction}
\end{figure}
The foregoing is a rather formal justification, and other more physical arguments
can be given (see, for example, Lenny Susskind's contribution to
these proceedings).   One way that one can see this rather clearly is to
consider two parallel slices at different values of $r$, as depicted in
Figure \ref{rgaction}.   If one considers the correlator of two operators on 
a brane then the correlator is only sensitive to the region of the space-time
that is relatively near both operators.  If the operators are widely separated, as are  
${\cal O}_1$ and ${\cal O}_2$ in Figure \ref{rgaction},  then they sample
deeply into the interior of the space-time (down to lower values of $r$). 
 If the operators are close together like ${\cal O}_3$ and ${\cal O}_4$ then
they are only sensitive to details of the space-time at large $r$.  Thus
we see the UV-IR duality of holgraphy:  Short distance behaviour on the brane 
depends upon large $r$, while large distance behaviour on the brane 
depends upon small values of $r$.

One can take a more active view of the same process.  Consider a 
fixed volume, say $1 m^3$,  on the $D3$-brane measured in the 
Poincar\'e metric, but consider this volume at two different values of $r$.  
This is depicted in Figure \ref{rgaction} by the large squares on the left and 
right.  If one uses supergravity to evolve data from large $r$ to smaller
$r$ then it will be blue-shifted.  Thus if one uses supergravity equations
to determine evolution in $r$, studying physics on a Poincar\'e scale
of $1 m^3$ at smaller $r$ will involve averaging over data from several
regions of size $1 m^3$ at the larger value of $r$.  This is precisely 
Wilsonian coarse-graining and it comes out of holography
via gravitational red (blue)-shifts.  The evolution to smaller
values of $r$ is thus renormalization group flow to the infra-red.

\subsection{Central charge and cosmological entropy}

There are several ways in which one can put computational flesh on
these ideas, but one of the simplest and most beautiful is the
the holographic $c$-theorem. 

In a conformal field theory the central charge essentially counts the
number of massless degrees of freedom in the theory.  It may be defined
in terms of the leading singularity in the correlator, or operator
product, of two energy-momentum tensors.  In $d$ dimensions the
energy-momentum tensor, $T(z)$, has dimension $d$, and so to leading 
order one has
\begin{equation}
T(x) ~ T(y) ~\sim~ {C \over |x -y |^{2 \,d} } \,.
\end{equation}
(I am suppressing all the details of tensor indices.)
If $T(x)$ is cononically normalized then the constant $C$ is
essentially the central charge, and it is proportional to the 
number of degrees of freedom within the theory:
It simply counts the ways in which energy can be transmitted.   

In a holographic theory one can argue that the central charge
is dual to a power of the ``effective cosmological constant,''  that is, 
\begin{equation}
C(r)  ~\sim~ {1 \over A'(r)^{d-1}} \,,
\end{equation}
where $A(r)$ is the function in (\ref{tenmet}).  The central charge
thus depends upon the scale $r$.  Moreover, it can shown 
\cite{Freedman:1999gp} that
if the matter in the supergravity theory obeys a weak energy condition,
then $C(r)$ monotonically decreases as $r$ decreases, and is only
stationary in an AdS vacuum, and hence at a conformal fixed point on the brane.
In other words, the number of dynamical degrees of freedom monotonically 
decreases as one flows to the infra-red, unless the flow reaches a 
conformal fixed point and then the number of degrees of freedom remains
constant.  This is precisely what one should expect as a result of 
coarse-graining:  a gradual loss of non-scale invariant degrees of freedom.

In cosmology, where the evolution is over time, the quantity analogous to
$A'(r)^{d-1}$ is the entropy density of the universe, and this
is monotonically increasing.   Thus there is at least a formal link
between the entropy in cosmology and the central charge in holographic
renormalization group flows.  Moreover, both the flow of entropy and the 
flow of the central charge reflect a loss of information about the detailed
finer structure of the matter within the space-time.

\subsection{Universality and black branes}

The foregoing shows that there are remarkable links between classical results
of relativity and quantum properties of field theory on the brane.  Much has been
done to develop this,  but given that this meeting is to celebrate Stephen's
contributions to science, there is one speculation that seems very appropriate.

In quantum field theory the idea of universality loosely states that the (infra-red) 
end-point  of a  renormalization group flow does not depend upon ultra-violet details.
In low dimensions one can go further and argue that infra-red renormalization
group fixed points do not depend upon details of interactions, but merely depend
upon the symmetries of, and number of degrees of freedom in the physical system.
If one thinks holographically, then this says that the supergravity solution for small 
values of $r$ does not depend upon the details of the matter at large values of $r$.
This begins to sound like a ``no-hair''  theorem, particularly when one thinks 
of the stronger low-dimensional idea of universality.  If the flow solution evolves
(as many of them do) to a black brane in the core, then indeed universality of
the field theory on brane, and a ``no hair'' theorem for the black brane are trying
to capture exactly the same physical ideas.

It should, of course, be remembered that in holography of four-dimensional
field theories  we are interested in radial evolution in five, or even ten, dimensions.  
As Gary Horowitz's talk illustrates, the collapse of
black-branes in higher dimensions can lead to rather exotic  end-states.
There is, however, a very interesting convergence: One would like
to know all the infra-red fixed points of Yang-Mills theories, and it is
intriguing to think that (at least for large $N$) it might rest upon
understanding the possible end-states of collapse of black branes.

\section{Holgraphic renormalization group flows: An example}

One of the simplest ways to exhibit a holographic renormalization
group flow is to take a conformally invariant holographic theory
and perturb it by a relevant operator and then use supergravity to
study the flow.  One may also seek out fixed points of such a flow
by looking for new, non-trivial AdS vacua that might be approached
in the infra-red ($r \to - \infty$).  

Gauged ${\cal N}=8$ supergravity
in five dimensions provides a powerful tool for analyzing  a 
sub-class of flows in ${\cal N}=4$ Yang-Mills theory
\cite{Distler:1998gb,Girardello:1998pd,Freedman:1999gp}.  The 42 scalars
of the supergravity are dual to bilinear operators in the Yang-Mills theory,
and thus by choosing the appropriate supergravity scalar one
can introduce a gauge invariant mass term for any field in the
Yang-Mills theory.  The flow to the infra-red will then correspond to
integrating the massive field out of the Yang-Mills action, leaving
a reduced number of degrees of freedom, and perhaps some
new, non-trivial interactions.    In the supergravity theory the
scalars corresponding to the mass terms must be Poincar\'e 
invariant, but will depend upons $r$, and their evolution will be
determined by the supergravity equations of motion.  The 
potential in the supergravity will thus determine the flow, and
critical points of the potential will, in principle, determine 
non-trivial conformal fixed points of a flow.   Therefore,
the (rigidly-determined) supergravity potential represents
and characterizes the phase diagram and flows of ${\cal N}=4$ 
Yang-Mills  theory under mass perturbations.

\subsection{Stability and unitarity}

There are several subtleties involved in holographic duality, and
the identification of flows, and one of these involves the stability issue.  
There are several known critical points of the gauged ${\cal N}=8$
supergravity potential \cite{Khavaev:1998fb}, and several of them fail the 
Breitenlohner-Freedman stability condition (\ref{BFbound}).  They
are thus not even perturbatively stable supergravity vacua.

It turns out that the holographic correspondence  yields a direct 
relationship between the mass of small oscillations in the
supergravity and the dimension of conformal operators
on the brane \cite{Gubser:1998bc,Witten:1998qj}.  If a supergravity
scalar fails the Breitenlohner-Freedman condition then the 
corresponding operator has imaginary conformal dimension,
and the field theory cannot be unitary.  Moreover, if one carefully
examines a flow from the maximally symmetric critical point 
to an unstable critical point one can see that it appears to correspond 
to an infinite energy deformation involving a mass and a vacuum
expectation value for the same operator.

This raises the question as to what critical points and flows 
are physically sensible.  The general belief is that if the critical
point is stable then it is a physical vacuum for both the supergravity
and the theory on the brane, but there is a further issue.  It should always
be remembered that the supergravity description  is really only
valid at large $N$, and so it may be that a flow, or a critical point
might represent a ``large N pathology'' that may not be physical 
for finite $N$.    Fortunately string theory gives us an answer:  At
finite $N$ one must use the full string path integral, and so a 
flow solution in supergravity will be valid at finite $N$ if we
can demonstrate that it is a good string vacuum.  The latter is 
not so easy to do, but based upon experience in string 
compactification in the late 1980's we know that supergravity
vacua are usually good approximations to string vacua if
they are supersymmetric.  Such vacua also have the virtue
of being completely semi-classically stable.

We are thus led to the following proposal:  Any supersymmetric
supergravity flow solution will reflect a real physical flow at 
finite $N$ for the theory on the brane.  This proposal
has passed several very non-trivial tests, and I will now
outline one of them.

\subsection{A supersymmetric flow}

In ${\cal N}=4$ Yang-Mills theory there is one gauge
field, $A_\mu$, four fermions, $\lambda^a$, and six scalars,
$X^I$, all in the adjoint of $SU(N)$.  Consider the following
mass perturbation on the brane:
\begin{equation}
\Delta {\cal L} ~=~  m_1 ~{\rm Tr}~\big( \lambda^1~ \lambda^1\big) ~+~
m_2^2 ~{\rm Tr}~\big( (X^1)^2+  (X^2)^2\big) \,.
\label{masspert}
\end{equation}
If $m_1 = m_2$ then this perturbation preserves ${\cal N}=1$
supersymmetry on the brane.  Moreover, in field
theory at finite $N$, this particular flow is known to lead to
a non-trivial, ${\cal N}=1$ supersymmetric fixed point \cite{Leigh:1995ep}.

\begin{figure}[h]
\hbox{\epsfxsize=2.1in 
\epsfbox{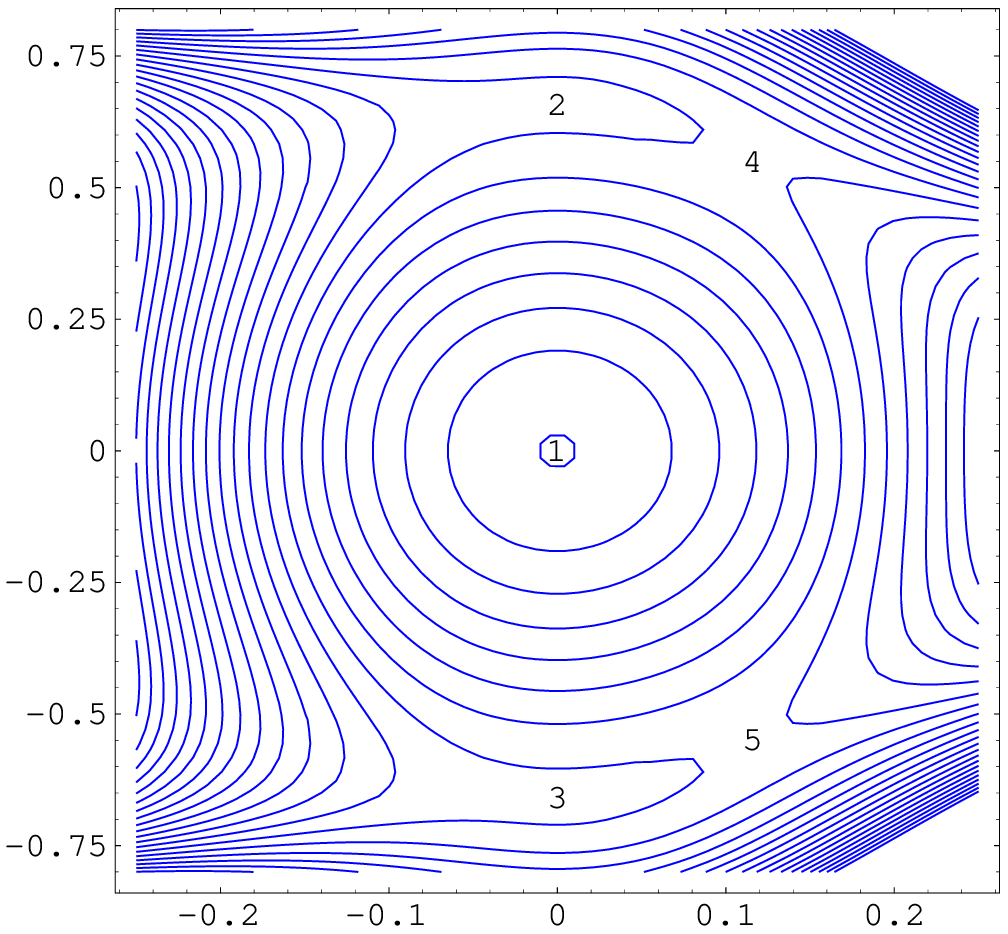} 
\hskip 0.205in
\epsfxsize= 2.1in  
\epsfbox{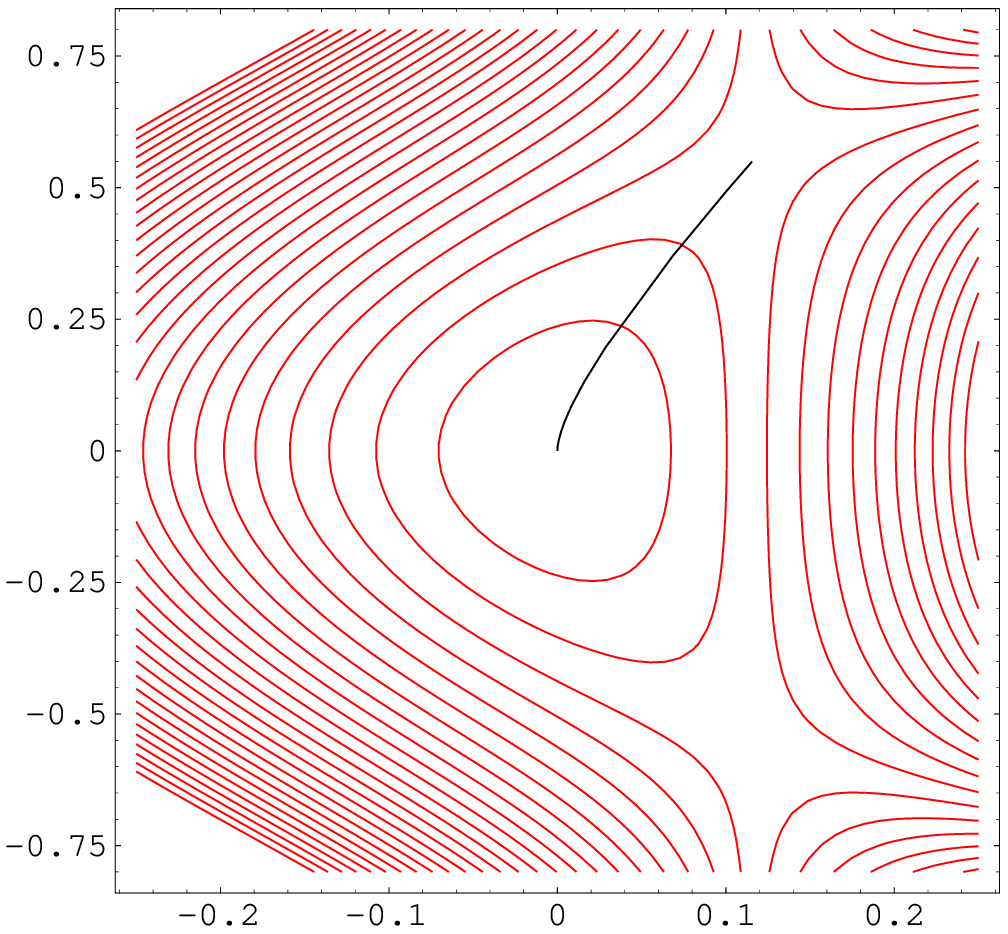}}
 \vspace{4.5pt}
\caption{The contour diagrams of part of the supergravity potential
and superpotential.
The mass parameters $\varphi_1 = m_2^2$ and $\varphi_2 = m_1$
are plotted along the horizontal and vertical axes respectively.  
Critical points are numbered on the plot of the potential, and the 
line on the second contour plot shows the steepest descent between
critical points of the superpotential. }
\label{contplots}
\end{figure}
In supergravity the perturbation (\ref{masspert})  is represented
by two scalars, $\varphi_1 = m_2^2$ and $\varphi_2 = m_1$.  On this
sub-sector the supergravity potential is easily computed, and it is
in fact shown in Figure \ref{Vplotfig}.  The contours of this potential are
shown in Figure \ref{contplots}. 

The supergravity equations of motion are, of course, second order
and involve the potential, ${\cal V}$.  However, if one seeks the
supersymmetric flow then this is given by solving a first-order system
involving a superpotential.  That is one solves:
\begin{equation}
 {d \varphi_j \over dr} ~=~{1 \over L}~{\partial {\cal W} \over \partial \varphi_j} \,,
\qquad {d A(r)  \over dr} ~=~- {2 \over 3 L} ~{\cal W}   \,,
\label{steepdesc}
\end{equation}
where $L$ is the AdS radius as $r \to \infty$, and 
\begin{equation}
 {\cal W} ~\equiv~   {1 \over 4 \rho^2}~ \Big[\cosh(2 \varphi_2)~
( \rho^{6}~-~ 2)~ - ( 3\rho^{6} ~+~ 2 ) \Big]  \,, \qquad \rho~\equiv~
 e^{{1 \over \sqrt{6}} \varphi_1} \,.
\label{Wdefn}
\end{equation}
The supergravity potential is then given by:
\begin{equation}
{\cal V} ~=~ {1 \over 2 \,L^2}~\sum_{j = 1}^2 ~\Big| {\partial {\cal W}
\over \partial \varphi_j} \Big|^2 ~-~ {4 \over 3 \,L^2}~\big|{\cal W} \big|^2 \,,
\label{VfromW}
\end{equation}

The contour diagrams in Figure \ref{contplots} show several
critical points.  The central one is the maximally supersymmetric
vacuum in which all the supergravity scalars vanish.  The other
vacua come in pairs related by a trivial reflection.  The vacua labelled
2 and 3 are unstable, while the vacua labelled by 4 and 5 are 
${\cal N}=2$ supersymmetric in the bulk (${\cal N}=1$ supersymmetric 
on the brane) and preserve $SU(2) \times U(1) \subset SO(6)$.
Only the supersymmetric vacua show up on the contour plot
of the superpotential, ${\cal W}$.
Thus there is  a good candidate supergravity vacuum state for
the field theory fixed point of \cite{Leigh:1995ep}.  Indeed it
is easy to verify that the unbroken supersymmetry and 
$R$-symmetry at the non-trivial critical point exactly matches 
that found in \cite{Leigh:1995ep}.

Equations (\ref{steepdesc}) show that the flow is a
given by steepest descent on ${\cal W}$, and that the cosmological
function, $A(r)$, is completely determined by the steepest descent.
This steepest descent is shown in Figure \ref{contplots}.  One can
see that near the central critical point one has 
 $\varphi_1  \sim \varphi_2^2$, which is consistent with $m_1 =m_2$
as required by supersymmetry.   At the non-trivial critical
point one can examine the linearized supergravity spectrum and
one finds that it matches perfectly with holographically dual
operators that one expects to find in the field theory \cite{Freedman:1999gp}.  
Most significantly, the supergravity predicts the following ratios of central charges:
\begin{equation}
 {C_{IR}  \over C_{IR} }~=~ \bigg( {A'( - \infty)  \over A'( + 
\infty)}\bigg)^3 ~=~   {27 \over 32 }  \,,
\label{Cratio}
\end{equation}
This can be checked against a direct, and rather non-trivial,
anomaly computation within the field theory, and there is perfect agreement.

\subsection{An important open problem}

One does not have to find restrict one's attention to flows to non-trivial fixed 
points:  There are many physically interesting families of flows that approach 
singular metrics at finite values of $r$. Among the most interesting of these
are the half-maximal supersymmetric flows (${\cal N}=2$ on the brane
 ${\cal N}=4$  in the bulk).  These are particularly important because
part of the quantum effective action is exactly 
known \cite{Seiberg:1994rs,Seiberg:1994aj}, and thus one can
perform some extremely non-trivial tests of such holographic flows.

\begin{figure}[h]
\hbox{\epsfxsize=2.5in 
\epsfbox{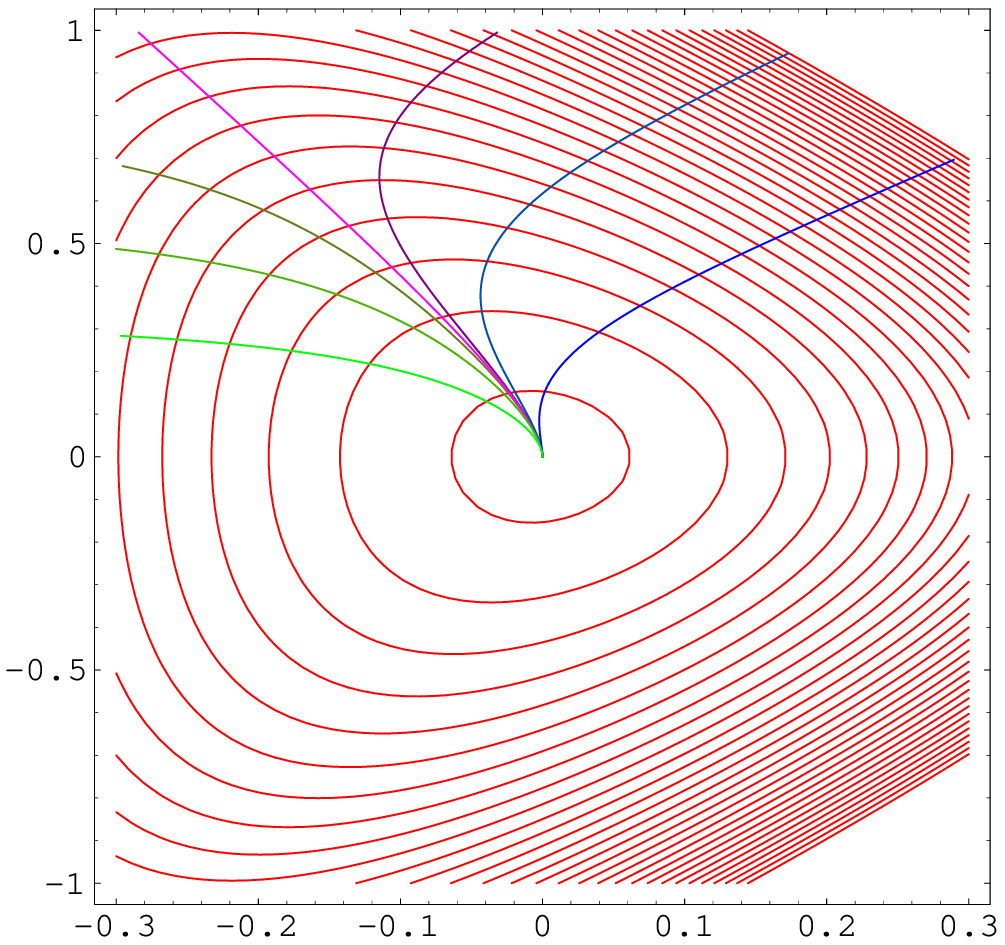} 
\hskip 0.3in
\epsfxsize= 1.8in  
\epsfbox{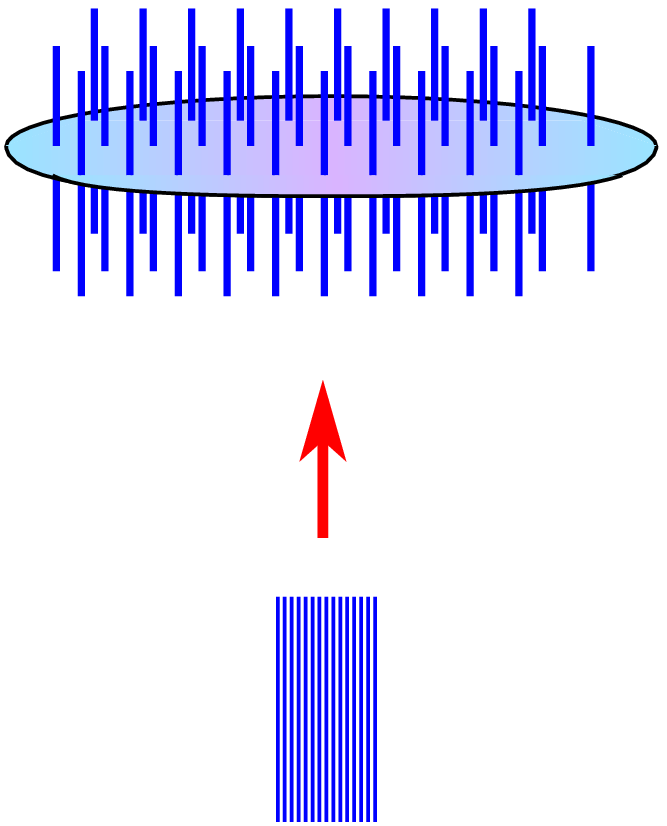}}
 \vspace{4.5pt}
\caption{Half-maximal supersymmetric flows appear as
steepest descents on a supepergravity superpotential.
The physically most interesting flow follows the ridge line.
This flow is singular at finite $r$, and corresponds  to a 
``disk-like'' smearing out of the $D3$-brane
distribution.}
\label{SWflows}
\end{figure}
Rather surprisingly, only a little is known about this class of solutions 
in $IIB$ supergravity.  Indeed only one such flow solution is known,
and this was constructed again using the techniques of gauged
${\cal N}=8$ supergravity, and the result was then lifted to the full
ten-dimensional theory \cite{Pilch:2000ue}.  Physically one again looks
at steepest descents on another supergravity superpotential, as shown
in Figure \ref{SWflows}.    In the corresponding ten-dimensional
solution the branes are now located at finite $r$ and are spread apart
over a uniform ``disk-like'' distribution.  The asymptotic 
behavior of this solution can be studied in detail \cite{Buchel:2000cn}
and one does indeed find results that are beautifully consistent
with those of the quantum field theory.

The solution of  \cite{Pilch:2000ue} represents only one fixed
(in fact, the most uniform) distribution of smeared out $D3$-branes.
The quantum field theory results of  \cite{Seiberg:1994rs,Seiberg:1994aj}  
allow, and give exact results, for an arbitrary  two-dimensional distribution
of the $D3$-branes.  Thus the solution of \cite{Pilch:2000ue} is but a single
point in an infinite moduli space of half-maximally supersymmetric solutions.
No other solutions in this class are known.

For this reason, and for several others it would be very nice
to find general results for half-maximal supersymmetric backgrounds.
There are theorems about hyper-K\"ahler manifolds
leading to such backgrounds in the absence of branes, and so the open
question is  to find the analogue of the hyper-K\"ahler condition
for half-maximal supersymmetry in the presence of branes and
the Ramond-Ramond fluxes that they generate.

\section{Final Comments}

In the late 1980's an eminent particle physicist publicly
compared string theory to President Reagan's rather misguided
Strategic Defence Initiative, commonly called ``Star Wars.''   The suggestion
was that both programs were making extravagant claims based upon little, or
no experimental evidence.   While this commentary probably grew
out of some close-mindedness on one side and  a certain amount of
hubris on the other, it serves to underscore some much broader
issues that concern string theory and string theorists.

First, the problem of quantum gravity is very unusual in science:  the
usual methodology is that one develops theories and chooses between 
them based on experimental data.  In quantum gravity we have the
problem that there are fundamental inconsistencies between two classes
of well tested theory, and the problem has been to find {\it any}
theory that solves the most basic of theoretical constraints.  
Stephen's work on black hole radiance and upon information loss
has been extremely important in making us aware of some
central aspects of what happens when quantum mechanics meets
gravity.  In particle physics, supergravity was a very important
step in addressing the problem of divergences, and string theory is 
almost certainly finite.  String theory remains the {\it only}
viable theory of quantum gravity  that we have found after 50 years of 
trying.    This is a remarkable achievement, but it is has remained
rather esoteric, and perhaps not as widely appreciated
as it might, and should be.

This is further complicated by the fact that research in string theory does 
not fit the classical ``scientific method:''  Predict and then experimentally test.  
Instead we test string theory in as many  ways as we can computationally, 
and we study it in many, many limits.   What is remarkable is that when 
we do this we often get new and deeper insights not just into quantum gravity 
but into particle physics and mathematics:  String theory created  the 
field of mirror symmetry in algebraic geometry; it has given new methods 
of computing multi-gluon amplitudes in QCD;  it has spun-off a whole industry on 
``brane-worlds''   and, as other speakers at this meeting have described, string
theory and $D$-branes have given new insight into the
apparent information loss in black holes.  
In this talk  I have tried to describe how string theory  has 
given us a new way of looking at and analyzing strongly coupled quantum 
field  theories via holography.  There are beautiful dualities between 
renormalization group flows, coarse graining and cosmological redshifts
and entropy.  There are simple classical calculations from which
one obtains insight into the phase structure of strongly coupled 
quantum field theory.  This, and all the other developments,
 are not experimental confirmation, but  they
do represent major progress in theoretical understanding.  I believe that
over the last few years the physics community has begun to
appreciate the remarkable list of string theory spin-offs because there
currently seems to be a more  broadly based enthusiasm for string 
theory than there was in the 1980's.

This still leaves open the issue of experimental confirmation:  
It should be remembered that a  quantum theory of gravity only  
becomes important at energy scales  vastly above the reach of any  
conceivable particle accelerator.  So, to
a purist, the experimental confirmation of string theory many take
a very, very long time.   On the other hand, supersymmetry is one of the 
foundational components of string theory, and current experimental data slightly 
favours the minimal {\it supersymmetric} standard model.  
At this meeting Edward Witten  predicted that within the next decade  we will 
see  supersymmetry  at the LHC.  In terms of a solution to quantum gravity 
it is by no means essential that this happens, but I think it is 
very important to further strong support of string theory within the 
research community.   

Finally, it is worth remembering that for about a decade after they
were first invented, supersymmetric theories were  considered 
by many to be  mathematical freaks with particle spectra
that are obviously ridiculous. Today, the study of  supersymmetric field 
theories is not only respectable, 
but it lies at the core of  particle physics phenomenology.  In 1980  it was not 
very clear what the future of supersymmetry would be, but it was obviously
a very important, new theoretical idea.  Stephen has always had a very
good sense for the important issues and the right questions to ask. 
I am therefore very grateful  to Stephen for his interest,
support and encouragement in the pursuit of a body of ideas that
has grown into one of the most exciting theoretical fields of the last, and
hopefully the next, twenty years.

\end{document}